# *In-vitro* Anti-bacterial Activity of Methanol and Aqueous Crude Extracts of *Horsfieldia iryaghedhi*

## RMHKK Rajapaksha [a], EMN Fernando [b], AWMKK Bandara [a], NRM Nelumdeniya [b] and ARN Silva [a*]

[a] *Department of Basic Sciences, Faculty of Allied Health Sciences, General Sir John Kotelawala Defence University, Werahera, Sri Lanka.*
[b] *Department of Pharmacy, Faculty of Allied Health Sciences, General Sir John Kotelawala Defence University, Sri Lanka.*

*Authors' contributions*

This work was carried out in collaboration among all authors. All authors read and approved the final manuscript.

## ABSTRACT

**Aims:** Over the past two decades, the rise of multidrug resistance (MDR) in bacteria has posed a significant threat to global health. The urgent need for new treatment alternatives has brought attention to the potential of plants, which harbor a wealth of unexplored phytochemicals with therapeutic properties. This study aims to evaluate the anti-bacterial efficacy of methanol and aqueous extracts from the leaves and bark of *Horsfieldia iryaghedhi In vitro*.
**Methodology:** Aqueous and methanol extracts were obtained from the cold maceration method. *In vitro* anti-bacterial activity of methanol and aqueous leaf, bark, and combination extracts were determined against gram-negative bacteria *Escherichia coli* (ATCC® 25922) and gram-positive bacteria *Staphylococcus aureus* (ATCC® 25923). The anti-bacterial assay for different concentrations of each extract was conducted through the well-diffusion method, with Gentamycin serving as the positive control.

---

*\*Corresponding author: E-mail: nsrajith@kdu.ac.lk; nsrajith2005@yahoo.com;*

**Results:** Methanol leaf and combination extracts of *Horsfieldia iryaghedhi* have shown a positive anti-bacterial response at their highest concentrations of 1000μg/mL and 500μg/mL against gram-positive bacteria *Staphylococcus aureus* while none of the extracts showed anti-bacterial activity against gram-negative *E. coli* at the experimented concentrations.

**Conclusion:** The study concludes that methanol extracts of *H.iryaghedhi* should be further analyzed for their anti-bacterial activity, and there could be potential lead molecules that can be developed as antibiotics.

## 1. INTRODUCTION

Antibiotics have played a crucial role in saving humanity from the debilitating effects of numerous infectious diseases caused by bacteria. Despite their continued usefulness, overuse, and considerable misuse have led to resistance among many infectious organisms, creating an urgent global need for alternative solutions. Factors such as an unfavorable regulatory climate, scientific challenges, low financial returns, and industry consolidation have prompted many pharmaceutical companies to withdraw from antibiotic drug development and antibiotic discovery is not keeping pace with the escalating levels of antibiotic resistance, a pressing global health concern. Hence, It is crucial to increase the productivity of the antibiotic drug discovery pipeline to address the current crisis and deliver drugs for challenging infectious diseases. Screening programs have been initiated to address these challenges and natural product research still holds promise as a source of new molecules for drug discovery [1]. Herbal medicines have been acknowledged for their effectiveness, minimal side effects, and affordability compared to Western medicines, and many medicinal plants are valuable natural resources that have been utilized in the development of novel drugs [2]. According to the World Health Organization (WHO), over 80000 species of higher plants have been documented for their therapeutic use out of the world's 25 million species. Furthermore, it is estimated that roughly 21000 plant species have the potential to be utilized as therapeutic agents [3].

The Myristicaceae family encompasses diverse flowering plants native to various regions, including Africa, Asia, the Pacific islands, and America. The family is also known as the nutmeg family, due to its most renowned member, *Myristica fragrans*, which is the source of nutmeg. The family boasts a remarkable diversity, comprising approximately 520 species of trees and shrubs found in tropical forests worldwide. These plants are characterized by their fragrant nature and distinctive aroma that hold significant medicinal value. Myristicaceae plants are known to address ailments such as stomach ulcers, indigestion, and liver disease. Additionally, they serve as an emmenagogue (stimulating menstrual flow), a nerve tonic, a diuretic, and a diaphoretic (promoting sweating). Furthermore, these botanicals are considered aphrodisiacs. Notably, essential oils extracted from certain genera within the Myristicaceae family, such as Virola, exhibit anti-fungal and anti-bacterial properties. The dark red resin from the tree bark of Virola contains a mix of hallucinogenic alkaloids [4].

*Horsfieldia iryaghedhi*, also known as the "Malaboda" tree or "ruk", is a fast-growing flowering plant that is native to Sri Lanka and typically attains heights ranging from 10 to 20 meters and thrives in Sri Lanka's wet zone, particularly along the borders of paddy fields, water streams, and rivers. It is a large tree with a straight, tall trunk and many long drooping branches. It has a thin bark that is brownish-grey, and slightly cracked. Branchlets are marked with leaf scars, and young parts have orange tomentose. The plant consists of large leaves that are simple, alternate, and spread distichously. Leaves are 17.5-30cm long and are oblong-lanceolate. They are acute or slightly rounded at the base, acuminate, bright green, and look polished on the upper surface. They have orange stellate tomentum beneath. Its petioles are 2.5cm long, very stout, and have rufous tomentose. *H. iryaghedhi* has regular flowers which are unisexual and dioecious. Numerous male flowers are present, which are orange-yellow and have a pleasant fragrance [4] Figs. 1 a) and b) show the flowers, leaves, and stems of *H. iryaghedhi*.

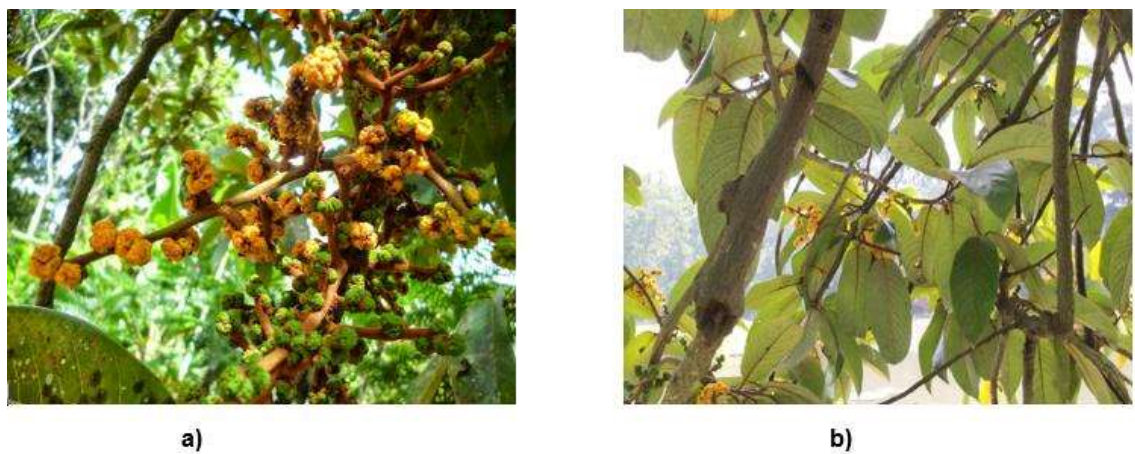

**Fig. 1. a) Flowers and b) Leaves and stem of the plant *Horsfieldia iryaghedhi***

Over 80 species of the genus *Horsfieldia* make up the Myristicaceae family, and most of these plants are employed extensively as antimicrobials in dermatological and cosmetic products [5]. Despite its prevalent use, the antimicrobial properties of the native *H. iryaghedhi* is remains unverified by scientific research. Hence, an attempt was made to evaluate the invitro antibacterial effect of crude aqueous and methanolic extracts of *H. iryaghedhi* on *Escherichia coli* and *Staphylococcus aureus* that were taken as representatives of gram-negative and gram-positive bacteria.

## 2. METHODOLOGY

### 2.1 Sample Collection

Fully expanded bark and leaves of *H. iryaghedhi*, each weighing 1500g, were collected in fresh conditions during daytime from Mattegoda Osu Uyana, Colombo District in Western Province of Sri Lanka (Coordinates: 6°48'05.4"N, 79°58'37.6"E).

### 2.2 Identification and Authentication of the Plant

For authentication, the adequately dried and pressed plant specimen was submitted to the Pharmaceutical Botany division at Ayurveda Research Institute, Nawinna, Sri Lanka.

### 2.3 Preparation of Crude Plant Material Extracts

Selected plant materials of *H. iryaghedhi* were thoroughly cleaned using running tap water and dried until a consistent weight was achieved. The dried plant materials were then powdered, and coarse powder samples were extracted.

#### 2.3.1 Methanol extraction procedure

Plant material powders, comprising 20g each of leaves, bark, and a combined mixture, were individually immersed in 160 ml of 99.9% methanol in separate closed glass bottles and were kept for seven days with occasional shaking in an orbital shaker. Subsequently, the extracts underwent filtration, initially through a dual-layer muslin cloth and then via Whatman No. 1 filter paper. The filtrate was concentrated to a dry residue using a rotary evaporator (HAHNSHIN Scientific-model no. H-2005V, SR no: V-00449), keeping the water bath temperature at 40°C. Finally, the residue was desiccated in ambient room temperature for two hours [6].

#### 2.3.2 Aqueous extraction procedure

Plant material powders, comprising 20g each of leaves, bark, and a combined mixture, were individually immersed in 160 ml of distilled water in separate closed glass bottles and were kept for 7 days with occasional shaking in an orbital shaker. Subsequently, the extracts underwent filtration, initially through a dual-layer muslin cloth and then via Whatman No. 1 filter paper. The filtrate was concentrated to a dry residue using a rotary evaporator (HAHNSHIN Scientific-model no. H-2005V, SR no: V-00449), keeping the water bath temperature at 60°C. Finally, the residue was desiccated in ambient room temperature for two hours [7].

## 2.4 Evaluation of *In vitro* Anti-bacterial Activity

### 2.4.1 Collection and sub-culturing of test microorganisms

Bacterial strains of *Escherichia coli* (ATCC® 25922) and *Staphylococcus aureus* (ATCC®25923) were received from the Medical Research Institute (MRI), Sri Lanka. These strains were sub-cultured on Mueller-Hinton agar plates and carefully maintained at 2-4°C for further studies [8].

### 2.4.2 Preparation of test solutions

Individual stock solutions for each plant extract variant (methanol leaf, methanol bark, methanol combination, aqueous leaf, aqueous bark, and aqueous combination) were prepared by dissolving 0.25g of the dry residue in 50 mL of distilled water. From these stock solutions, 4 mL was diluted with 20 mL of distilled water to achieve a concentration of 1000µg/mL. This dilution process was extended to create a series of test solution concentrations at 1000µg/mL, 500µg/mL, 250µg/mL, and 125µg/mL. [9].

### 2.4.3 Preparation of gentamycin antibiotic for positive control

Commercially available Gentamycin 40 mg/mL IV injection vial was used as the positive control. The standard solution was prepared by dissolving 0.5mL of Gentamycin in 20ml of distilled water to obtain a solution concentration of 1000µg/mL, and the concentration series for the standard solution was prepared as 1000µg/mL, 500µg/mL, 250µg/mL, and 125µg/mL [10].

### 2.4.4 Preparation mcfarland standards

A 0.5 Mcfarland standard was prepared in the laboratory by mixing 0.5mL aliquot of BaCl2 with 99.5 mL of H2SO4 with constant stirring to maintain a suspension [11].

### 2.4.5 Preparation of bacterial broth

Two to three bacterial colonies were carefully picked from subcultures using a sterile inoculating loop and transferred into 25 mL of 0.9% normal saline under aseptic conditions. The resulting bacterial suspensions were then evaluated against a Wickerham card to match the turbidity of the previously prepared McFarland Turbidity Standard. Adjustments were made to align the turbidity of the bacterial suspension with the 0.5 McFarland Standard, corresponding to a bacterial density of approximately $1.5 \times 10^8$ CFU/mL [12].

### 2.4.6 Preparation of culture media

19g of commercially available Mueller-Hinton agar was measured using an analytical balance and transferred to a 500 mL conical flask. 500 mL of distilled water was added, and the solution was gradually heated and stirred using a magnetic stirrer. Culture media was then autoclaved using an autoclave machine (TOMYKOGYO co. ltd, Model-SX-500, SR no; 49133064) for 15 minutes at 121° C temperature under 15 lbs pressure. Additionally, all the glass wear needed was kept in a hot air oven (BOV-V2225F with RS485) at 121°C temp for 2 hours [13].

### 2.4.7 Screening for anti-microbial activity using agar well diffusion method

11.11 mL of each bacterial suspension was aseptically introduced into 500 mL flasks containing Mueller-Hinton agar. A foundational layer of Mueller-Hinton agar was first poured and allowed to solidify. Subsequently, five sterilized aluminum cylinders, each measuring 8.0mm×6.0mm×6.0mm with an inner diameter of 6.0mm and open at both ends, were positioned on the solid agar layer using a template to ensure uniform spacing. The bacterial culture media was then homogenized and dispensed into a 130mm sterilized Petri dish under aseptic conditions, which was then left to cool and solidify. Finally, the cylinders were carefully removed using sterilized forceps [14]. All steps were executed under stringent aseptic conditions.

Each prepared plate, featuring five wells, was loaded with 200µl of the plant extracts at four different concentrations (1000µg/mL, 500µg/mL, 250µg/mL, and 125µg/mL), along with a positive and a negative controls.

All the plates were then incubated (CLW 240 IG SMART) at 37°C for 24 hours. Post-incubation, the diameters of the inhibition zones were measured with a Vernier caliper. The measurements from three independent trials were averaged to determine the mean zone of inhibition [15].

### 2.4.8 Determination of IC50 values

The collected data was statistically analyzed using SPSS version 23, and dose-response curves were plotted with GraphPad Prism 8.0.1 software. The data were analyzed using an independent sample t-test, and the results were considered statistically significant when the P-value was less than 0.05 [16].

## 3. RESULTS AND DISCUSSION

Plants have long been recognized as a valuable source of compounds for treating infectious diseases. They produce complex suites of compounds known as secondary metabolites that allow them to defend against pathogens which include phenol, polyphenols, terpenoids, essential oils, flavonoids, and alkaloids that can interfere with bacteria through several mechanisms. These compounds also have the potential to offer protection to humans against infectious diseases. Naturally occurring plant-based compounds are increasingly being explored as an alternative or complementary approach to antibiotics.

Several kinds of secondary metabolites, including lignans, flavones, sterols, alkaloids, and essential oils, in the *H. iryaghedhi* give various therapeutic effects. However, limited investigations have been done on discovering the phytoconstituents of *H. iryaghedhi*, and there are few publications on its anti-bacterial activity. Further, these investigations have not examined the link between efficacy, potency, and dose-response. The current work examined the *In vitro* anti-bacterial activity and dose-response of aqueous and methanol extracts of the leaves, bark, and combination of the Sri Lankan endemic *H. iryaghedhi*.

The natural concentration-dependent bacterial killing of Gentamycin was observed in *Staphylococcus aureus* and *Escherichia coli* samples. Table 1 shows the Anti-bacterial effect of Gentamycin against *Staphylococcus aureus* and *Escherichia coli,* indicating the zone of inhibitions (mm), and Fig. 2 shows the dose-response curves.

Only the methanol leaf extract of *H. iryaghedhi* demonstrated anti-bacterial activity against *S.aureus* at 1000µg/mL and 500µg/mL concentrations. The zone of inhibition for 1000µg/mL and 500µg/mL concentrations were 19.6mm and 12.6mm, respectively, while the zone of inhibition for methanol combination (Leaf/bark) extracts of *H.iryaghedhi* were 13.25mm and 12.32mm for 1000µg/mL and 500µg/mL concentrations, respectively (see Table 2). Fig. 3 shows the Inhibition zones of a) Gentamycin b) *H. horsfieldia* methanolic leaf extract c) *H. horsfieldia* methanolic bark extract and d) *H. horsfieldia* methanolic leaf/bark extract combination against *S.aureus*.

Zero zone inhibitions were seen in all leaf, bark, and combination aqueous extracts of *H. iryaghedhi* against *S. aureus*. Further, no inhibition was seen for all (leaf, bark, and combination) methanol and aqueous extracts of *H. iryaghedhi* against *E. coli*. See Fig. 4 a) and b).

The study results showed that the methanol leaf and combination extracts of *H. iryaghedhi* demonstrated a positive anti-bacterial response at the highest concentrations tested against sample gram-positive bacteria *Staphylococcus aureus* (ATCC® 25923). Additionally, the study's findings indicated that *H. iryaghedhi* plant extracts showed no anti-bacterial activity against representative gram-negative bacteria, *E. coli*, at the experimented concentrations.

**Table 1. Anti-bacterial effect of Gentamycin against *Staphylococcus aureus* and *Escherichia coli***

| Concentration (µg/mL) | Zone of inhibition (mm) *Staphylococcus aureus* | Zone of inhibition (mm) *Escherichia coli* |
| --- | --- | --- |
| 125 | 16.0 ±0.333 | 12.00±0.577 |
| 250 | 23.5 ±0.333 | 13.50±0.167 |
| 500 | 28.0 ±0.577 | 15.25±0.289 |
| 1000 | 31.5 ±0.667 | 25.00±0.333 |

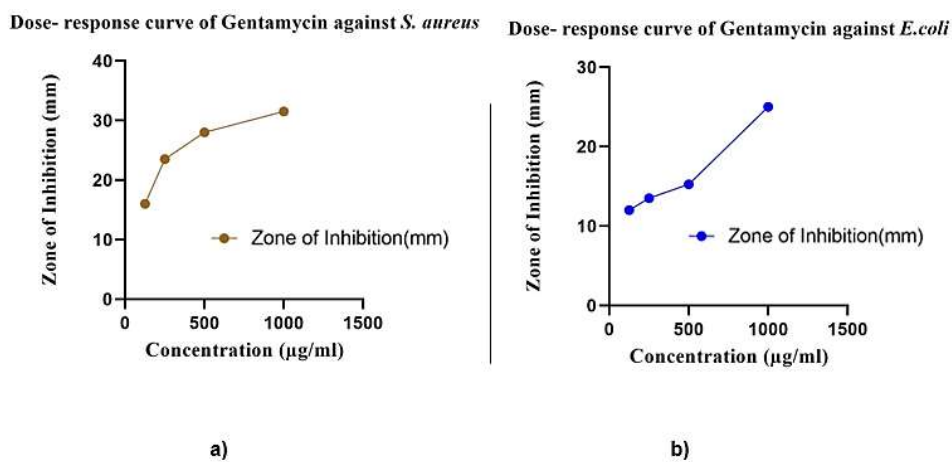

a)  b)

**Fig. 2. Dose-response curve of Gentamycin againt a)** *S.aureus* **b)** *E.coli*

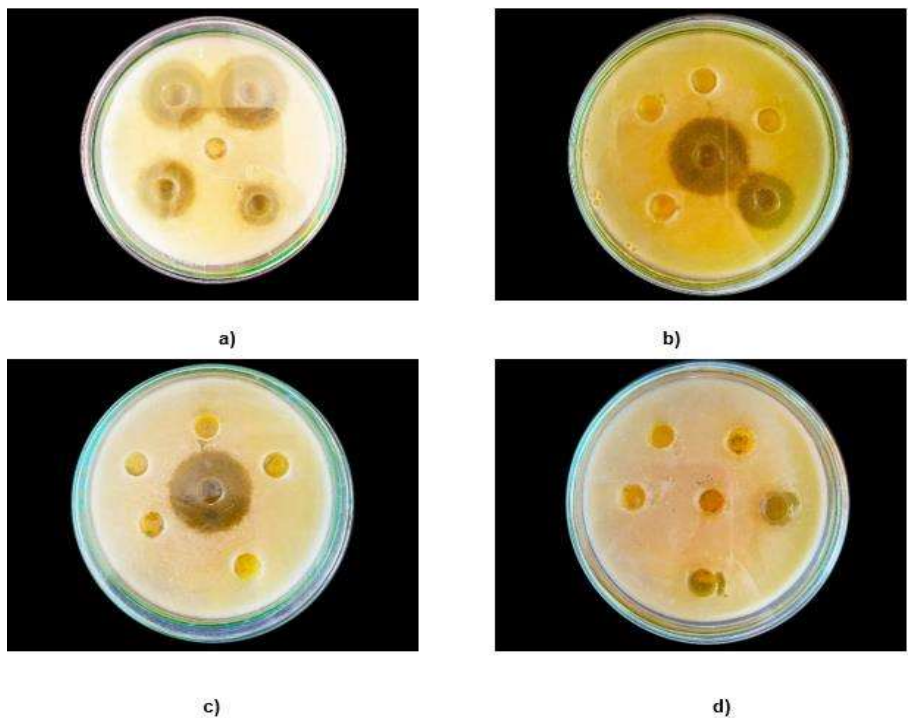

**Fig. 3. Inhibition zone of a) Gentamycin b)** *H.iryaghedhi* **methanolic leaf extract c)** *H.iryaghedhi* **methanolic bark extract, and d)** *H.iryaghedhi* **methanolic leaf/bark extract combination against** *S.aureus*

**Table 2. Anti-bacterial effect of methanol extracts of *H.iryaghedhi* against *S. aureus***

| Concentration(μg/mL) | Zone of inhibition (mm) | | |
|---|---|---|---|
| | Leaf | Bark | Combination |
| **125** | - | - | - |
| **250** | - | - | - |
| **500** | 12.6 ±0.667 | - | 12.32 ±0.667 |
| **1000** | 19.6 ±0.882 | - | 13.25 ±0.882 |

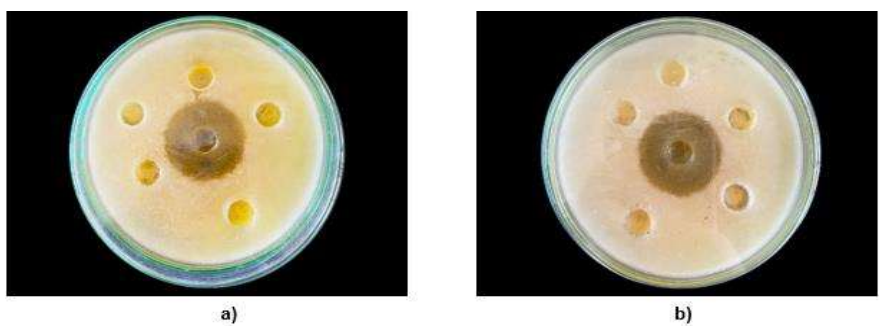

**Fig. 4. Effect of a) methanol extracts and b) aqueous extracts of *H.iryaghedhi* against *E.coli***

## 4. CONCLUSION

The present study highlights the anti-bacterial activity of methanol leaf extract and combined leaf and bark extracts of *H. iryaghedhi* against gram-positive bacterium *S. aureus* (ATCC® 25923) at their highest concentrations. However, the experimented aqueous extract concentrations of *H. iryaghedhi* plant did not exhibit anti-bacterial activity against the gram-negative bacterium, *E. coli*. This discovery opens avenues for in-depth exploration into the medicinal prospects of *H. iryaghedhi* plant extracts, specifically in isolating phytochemicals that could serve as lead compounds in antibiotic development.

## DISCLAIMER (ARTIFICIAL INTELLIGENCE)

Author(s) hereby declare that NO generative AI technologies such as Large Language Models (ChatGPT, COPILOT, etc) and text-to-image generators have been used during writing or editing of manuscripts.

## ACKNOWLEDGEMENTS

The authors would like to thank all the laboratory staff members of the Department of Pharmacy and Department of Basic Sciences, General Sir John Kotelawala Defence University, Werahera, Sri Lanka.

## COMPETING INTERESTS

The authors have declared that no competing interests exist.

*Peer-review history:*
*The peer review history for this paper can be accessed here:*
*https://www.sdiarticle5.com/review-history/116712*